\documentclass[]{aa}
\usepackage{graphics,latexsym,psfig}
\usepackage{graphicx}
\usepackage{longtable}

\def\asec{\ifmmode ^{\prime\prime}\else$^{\prime\prime}$\fi}
\def\amin{\ifmmode ^{\prime}\else$^{\prime}$\fi}
\def\degs{\ifmmode ^{\circ}\else$^{\circ}$\fi}
\def\etal{{et\,al. }}

\begin{document}

\title{The 1$^{st}$ INTEGRAL SPI-ACS Gamma-Ray Burst Catalogue}

\titlerunning{The 1$^{st}$ INTEGRAL SPI-ACS Gamma-Ray Burst Catalogue}

\author{
A. Rau \inst{1}        \and
A.v. Kienlin\inst{1}     \and
K. Hurley\inst{2}	\and
G.G. Lichti\inst{1}
}

\offprints{A. Rau, arau@mpe.mpg.de}

\institute{Max-Planck-Institut f\"ur extraterrestrische Physik,
  Postfach 1312, 85741
  Garching, Germany
\and UC Berkeley Space Sciences Laboratory, Berkeley, CA 94720-7450, USA
}

% --------------------------------------------------------------------------
\date{Received March 30, 2005 / Accepted April 15, 2005}

\abstract{We present the sample  of gamma-ray bursts detected with the
  anti-coincidence  shield  ACS   of  the  spectrometer  SPI  on-board
  INTEGRAL for  the first 26.5\,months  of mission operation (up  to Jan
  2005).  SPI-ACS works  as a  nearly omnidirectional  gamma-ray burst
  detector  above   $\sim$80\,keV  but  lacks   spatial  and  spectral
  information.   In this  catalogue, the  properties derived  from the
  50\,ms   light  curves   (e.g.,   $T_{90}$,  $C_{max}$,   $C_{int}$,
  variability, $V/V_{max}$) are given  for each candidate burst in the
  sample.   A  strong  excess  of  very short  events  with  durations
  $<$0.25\,s is  found. This population  is shown to  be significantly
  different from the short- and long-duration burst sample by means of
  the  intensity distribution  and $V/V_{max}$  test and  is certainly
  connected with cosmic ray hits in  the detector.  A rate of 0.3 true
  gamma-ray bursts per day is observed.
}

\maketitle

% --------------------------------------------------------------------------
\section{Introduction}

Although  discovered more  than  three decades  ago (Klebesadel  \etal
1973),  the  phenomenon of  cosmic  gamma-ray  bursts  (GRB) is  still
challenging,  with  many  open  questions to  solve.   Therefore,  the
detection and investigation of GRBs is one of the important scientific
objectives   of  ESA's   {\it  International   Gamma-Ray  Astrophysics
  Laboratory} (INTEGRAL) mission.  INTEGRAL contributes to GRB science
in two ways. (i) For bursts which  occur in the field of view (FoV) of
the  spectrometer  SPI  and  of  the imager  IBIS,  INTEGRAL  provides
accurate positions ($\sim$2\,arcmin) for rapid ground- and space-based
follow-up  observations.   In  addition,  high-energy spectra  in  the
20\,keV--8\,MeV range  are recorded (see  Mereghetti \etal 2004  for a
recent summary).   (ii) The  anti-coincidence system of  SPI (SPI-ACS)
acts as a nearly omnidirectional  GRB detector in the $\sim$80\,keV to
$\sim$10\,MeV energy range (von Kienlin \etal 2003).

It has  long been known  that there are  two distinct classes  of GRBs
which  differ  observationally  in  duration and  spectral  properties
(Mazets et al.  1981; Norris et al. 1984; Dezalay  et al. 1992; Hurley
1992;  Kouveliotou  et  al.  1993;  Norris et  al.  2000).   This  was
quantified using  data from the Burst and  Transient Source Experiment
(BATSE) detectors  (Fishman \etal  1989) on-board NASA's  {\it Compton
Gamma-ray Observatory}  (CGRO).  The sample of more  than one thousand
bursts  included in  the 4$^{th}$  BATSE catalogue  displayed  a clear
bimodal duration  distribution, with one group  having short durations
($<$2\,s)  and  the  other   longer  durations  (seconds  to  minutes)
(Kouveliotou \etal\ 1993). Statistically,  the peak energy of the Band
spectrum  for short  bursts  (Band \etal\  1993),  $E_{p}$, is  higher
(typically $\sim$100--1000\,keV) than that of the long-duration events
(typically   $\sim$70--500\,keV)  (Fishman   \&  Meegan   1995).   The
sensitivity of the  anti-coincidence system of SPI to  the short/ hard
burst  population is nearly  unprecedented, which  raised expectations
before  the  launch   of  the  mission  that  new   insights  into  the
distribution of short bursts would be achieved.

In this  paper we present  the 1$^{st}$ catalogue of  gamma-ray bursts
detected by SPI-ACS.  We  describe the instrumentation and data (Sect.
2),  the sample  selection criterion (Sect.3) and  the sample
analysis (Sect.4). In Sect.5 the duration- and intensity distributions
of the sample are discussed.

% --------------------------------------------------------------------------
\section{Instrumentation \& Data}

\begin{figure}
  \centering
  \includegraphics[width=0.45\textwidth,angle=0]{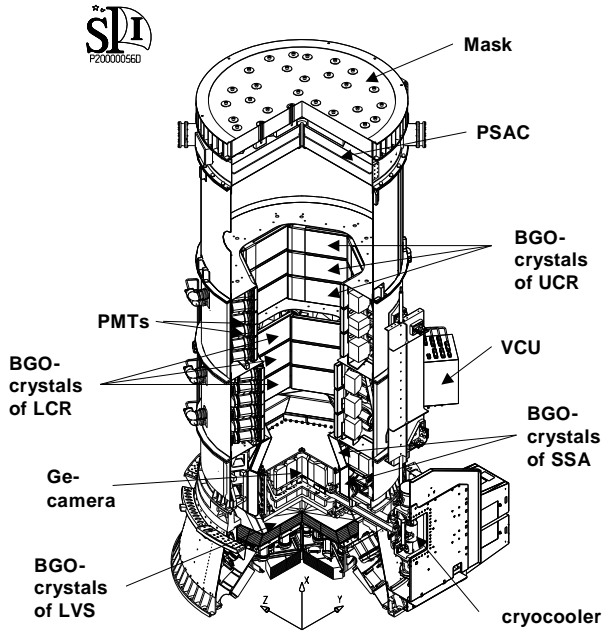}
  \caption{Spectrometer SPI and SPI-ACS. The SPI-ACS consists of the
  plastic scintillator (PSAC) and the BGO crystals of the upper (UCR)
  and lower collimator rings (LCR) plus the side-shield assembly (SSA)
  and lower veto shield (LVS). The crystals are monitored by the
  photomultiplier tubes (PMT) and the events are processed in the veto
  control unit (VCU).}
  \label{fig:spiacs} 
\end{figure}

The  anti-coincidence   system\footnote{developed  at  the  Max-Planck
Institute  for  extraterrestrial Physics,  Garching,  Germany} of  SPI
consists of 91  Bismuth Germanate (BGO) crystals with  a total mass of
512\,kg.  The thickness  of the individual crystals ranges  from 16 to
50\,mm.   They  are positioned  in  two  rings  (the upper  and  lower
collimator ring)  whose axes  are along the  viewing direction  of the
spectrometer, between the coded mask and the detector plane of SPI; in
addition,   there   are   side-shield   and   rear-shield   assemblies
(Fig.\ref{fig:spiacs}).   Incoming  $\gamma$-rays  are converted  into
optical  photons with  a  wavelength of  $\sim$480\,nm.   Each BGO  is
redundantly monitored  by two  photomultiplier tubes.  In  addition to
the BGO, a plastic scintillator (PSAC), whose purpose is the reduction
of the  511\,keV background produced  by particle interactions  in the
passive mask, is located directly below the coded mask.

%------------------------------------------------------------------
%------------------------------------------------------------------

Since the  ACS encloses the  SPI, it provides  a quasi-omnidirectional
field of view  with a large ($\sim$0.3\,m$^2$) effective  area for the
detection of gamma-rays (von Kienlin \etal 2003a).  Unfortunately, the
electronics does not allow for good spatial resolution; the burst data
consist only of  the total event rate from all  the crystals (for more
details see Lichti  \etal 2000) with a time  resolution of 50\,ms.  In
addition, however,  each BGO crystal is  read out every  96\,s with an
integration time of  1\,s, which in principle allows  a rough position
estimate for long and bright events.

The  SPI-ACS data  do not  include spectral  information  because the
energy of the gamma-rays is  not measured.  The only
information which  is avaiable is  that the energy of  the interacting
photon     must     be    above     the     threshold    energy     of
$\sim$80\,keV. Due to the different  properties of the
individual  crystals,   their  associated  photomultipliers,   and  the
redundancy concept (signals from two different crystals are  fed to the
same front-end  electronics) this energy  threshold can only be estimated
very coarsely.

As a  veto shield the SPI-ACS has  no upper limit to  the energy range
(von  Kienlin \etal  2003b).   Therefore, the  physical properties  of
individual  events  (e.g., peak  flux,  fluence)  cannot be  estimated
directly from the data.  A  conversion from detector counts to photons
cm$^{-2}$  s$^{-1}$   is  possible  only  for   bursts  whose  arrival
directions and spectra  (e.g., the Band function parameters  - Band et
al. 1993) are accurately  known from other gamma-ray instruments.  For
these bursts  the effective  area of the  detector can be  modelled by
simulations of the photon transport in the instrument as a function of
the incident  angle and  energy.  While the  number of GRBs  for which
this method can be applied is very small, it was especially useful for
the  determination of the  SPI-ACS spectral  parameters for  the giant
flare from SGR 1806-20 on Dec 27 2004 (Mereghetti \etal 2005).

SPI-ACS is part  of the INTEGRAL Burst Alert  System (IBAS; Mereghetti
\etal 2004).  A  software trigger algorithm searches for  an excess in
the overall count rate with  eight different time binnings (0.05, 0.1,
0.2,  0.4, 0.8,  1, 2,  \&  5\,s) with  respect to  a running  average
background. The trigger thresholds depend  on the time binning and are
set  to 9, 6,  9, 6,  9, 6,  9 \&  6$\sigma$, respectively.   For each
trigger an ASCII light curve (5\,s pre-trigger to 100\,s post-trigger)
and   the  spacecraft   ephemeris   are  stored   and  made   publicly
available\footnote{http://isdcarc.unige.ch/arc/FTP/ibas/spiacs/}.    It
was   originally   planned    to   distribute   transient   detections
automatically.  However, due to  the large number of spurious triggers
produced  by solar flares,  cosmic rays  and radiation  belt passages,
SPI-ACS alerts are now sent only manually to the interested community.
During the early mission phase many spurious triggers were produced by
rate increases in single 50\,ms time bins ($\sim$15 trigger per hour).
A software  solution was  found to remove  these events from  the data
stream.

Since  December 2002  SPI-ACS  has  been an  important  member of  the
3$^{rd}$  Interplanetary Network (IPN)  of gamma-ray  burst detectors,
which  provides  burst localizations  using  the triangulation  method
(Hurley 1997).  Thus the lack of spatial resolution of the ACS
can at least partly be compensated for.  By studying triangulations of a
number  of bursts  with  precisely known  localizations together  with
Konus/{\it  Wind}  and/or   Helicon/{\it  Coronas-F}  instruments  the
absolute timing of the instrument was adjusted and verified to an
uncertainty of 25\,ms (Rau \etal 2004).

%Burts with known fluences and peak fluxes (more than the number above)
%but without position and  spectral shape a cross-calibration with data
%from other  gamma-ray instruments is  limited by the diversity  of the
%spectral  properties, e.g.   peak energy  and spectral  slope.  As the
%observed fluxes  etc depend strongly  on the energy band  and incident
%angle,    a   calibration    like    that   will    have   a    strong
%scatter/uncertainty.  Thus,  we  will  focus on  the  measured  values
%(counts) for the  rest of the paper and leave  spectral results to the
%dedicated analysis of individual events.

% --------------------------------------------------------------------------
\section{Sample Selection}

A sophisticated statistical analysis of  the GRB events in the SPI-ACS
overall   count   rate  requires   a   thorough   and  robust   sample
selection. Generally, the best  method to distinguish gamma-ray bursts
from other events occurring in the data stream is their localization in
celestial coordinates. Likewise, detections of  the events by other missions with
gamma-ray detectors,  e.g. the  instruments participating in  the IPN,
can be used to discriminate  between real cosmic events and those
with a solar or particle origin.  Due to the lack of spatial information in the
SPI-ACS data,   observations with independent instruments are required in
order to apply this localization-based selection criterion.

Due to the different energy  ranges, sensitivities, and  parts of  the sky
observed by  the various currently active gamma-ray  detectors, the
above-mentioned  method of selection applies only to a sub-sample of
the   GRBs  observed  with   SPI-ACS.   Due  to   its  nearly
omnidirectional  field of  view and sensitivity, a  considerable fraction  of  GRBs is
expected to  be detected only by  SPI-ACS.  Thus, in order  to study a
statistically  more  significant  sample,  we  defined  our  selection
criteria to be independent of  the  confirmation by  other instruments  and
based only on  the  SPI-ACS
light curve.

\begin{figure*}
  \centering
   \includegraphics[width=0.31\textwidth,angle=270]{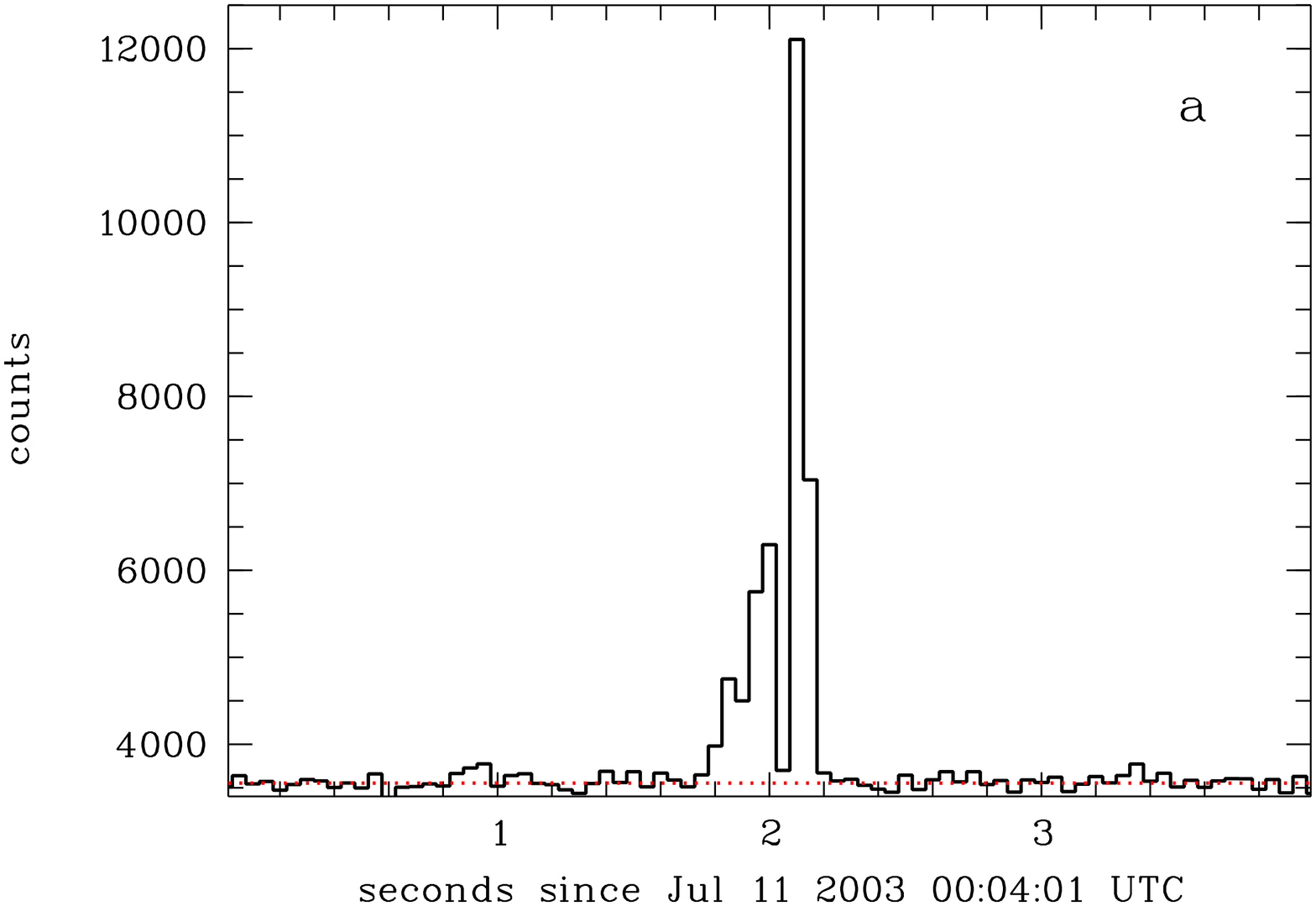}
   \includegraphics[width=0.31\textwidth,angle=270]{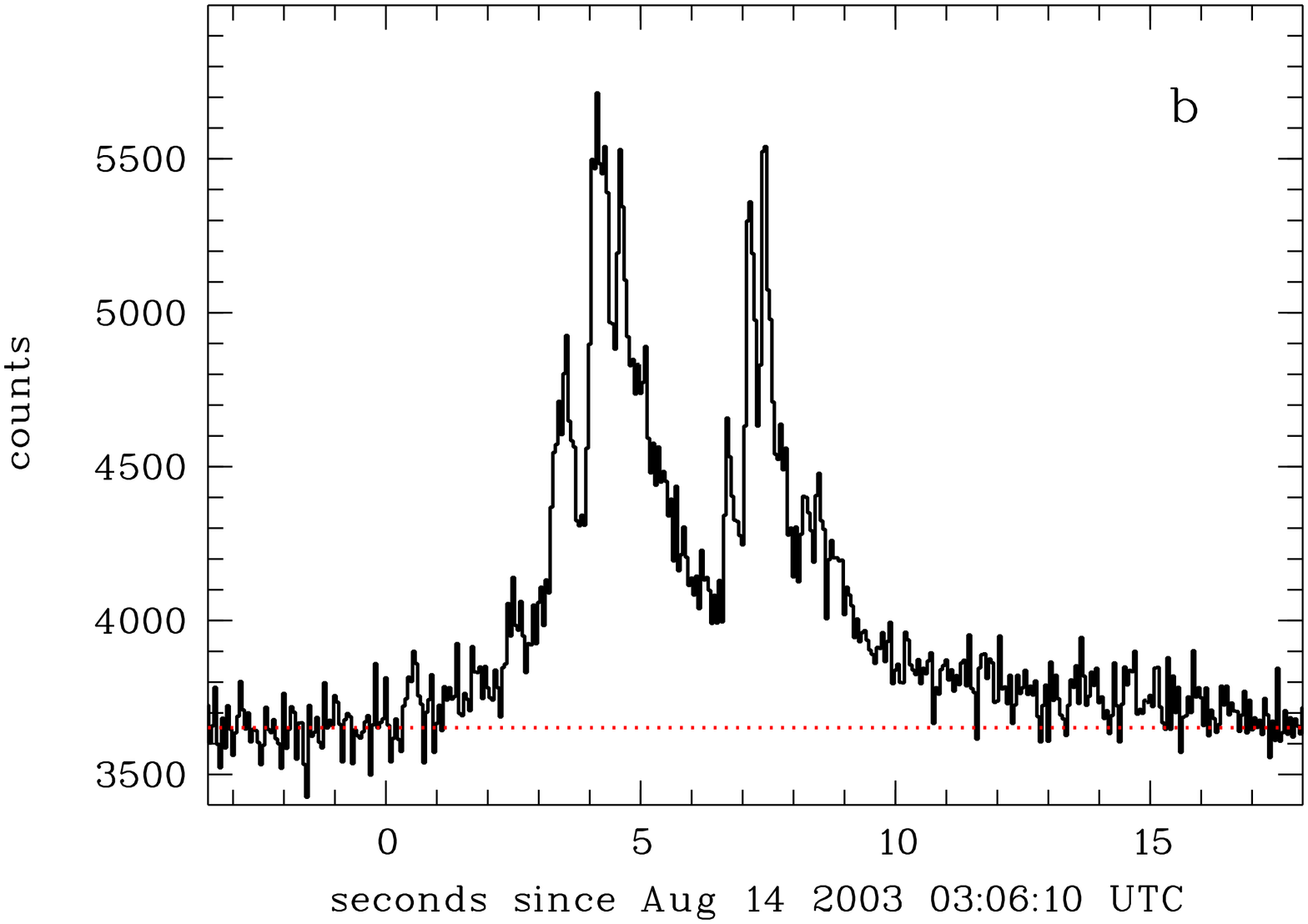}
   \includegraphics[width=0.31\textwidth,angle=270]{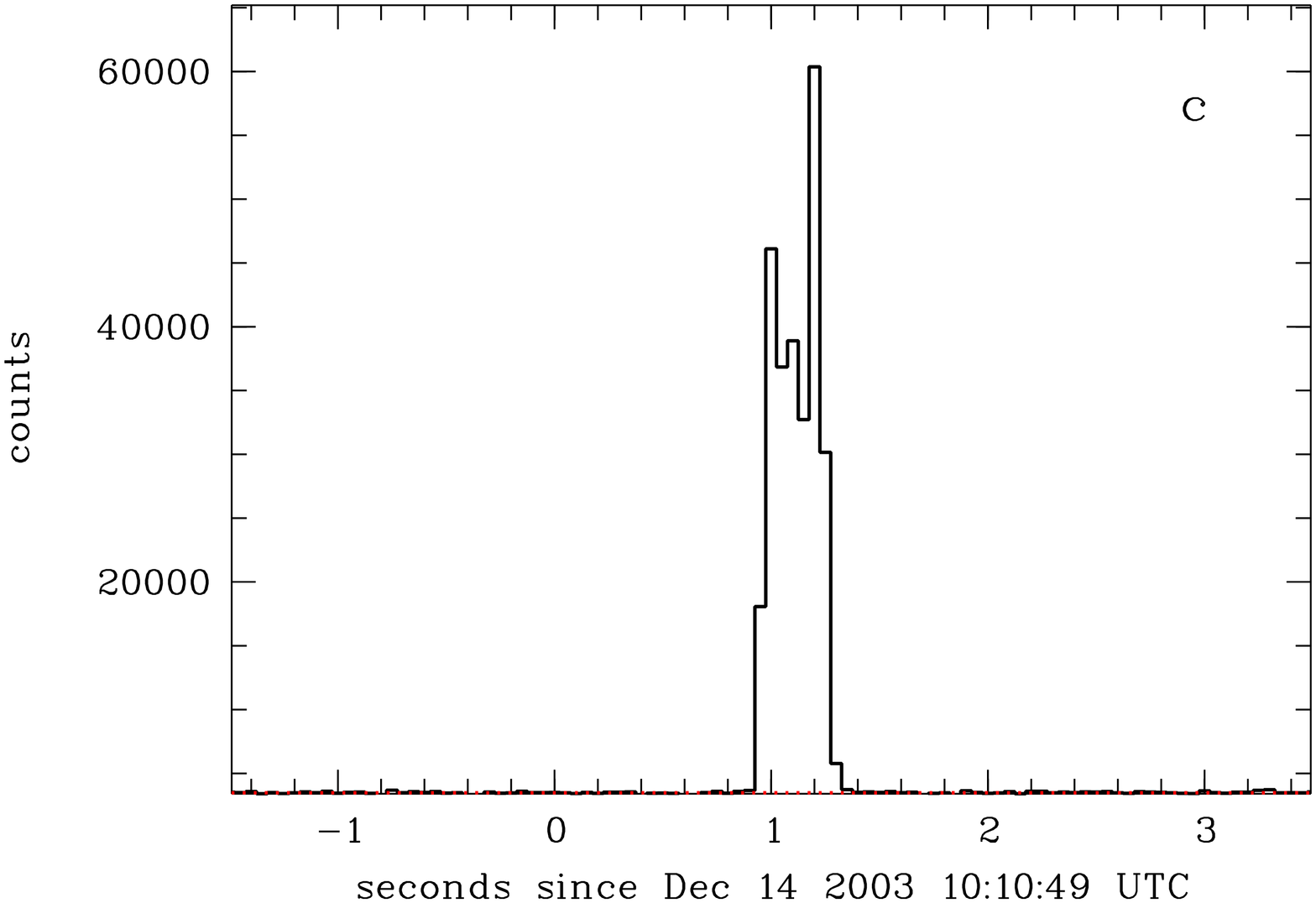}
   \includegraphics[width=0.31\textwidth,angle=270]{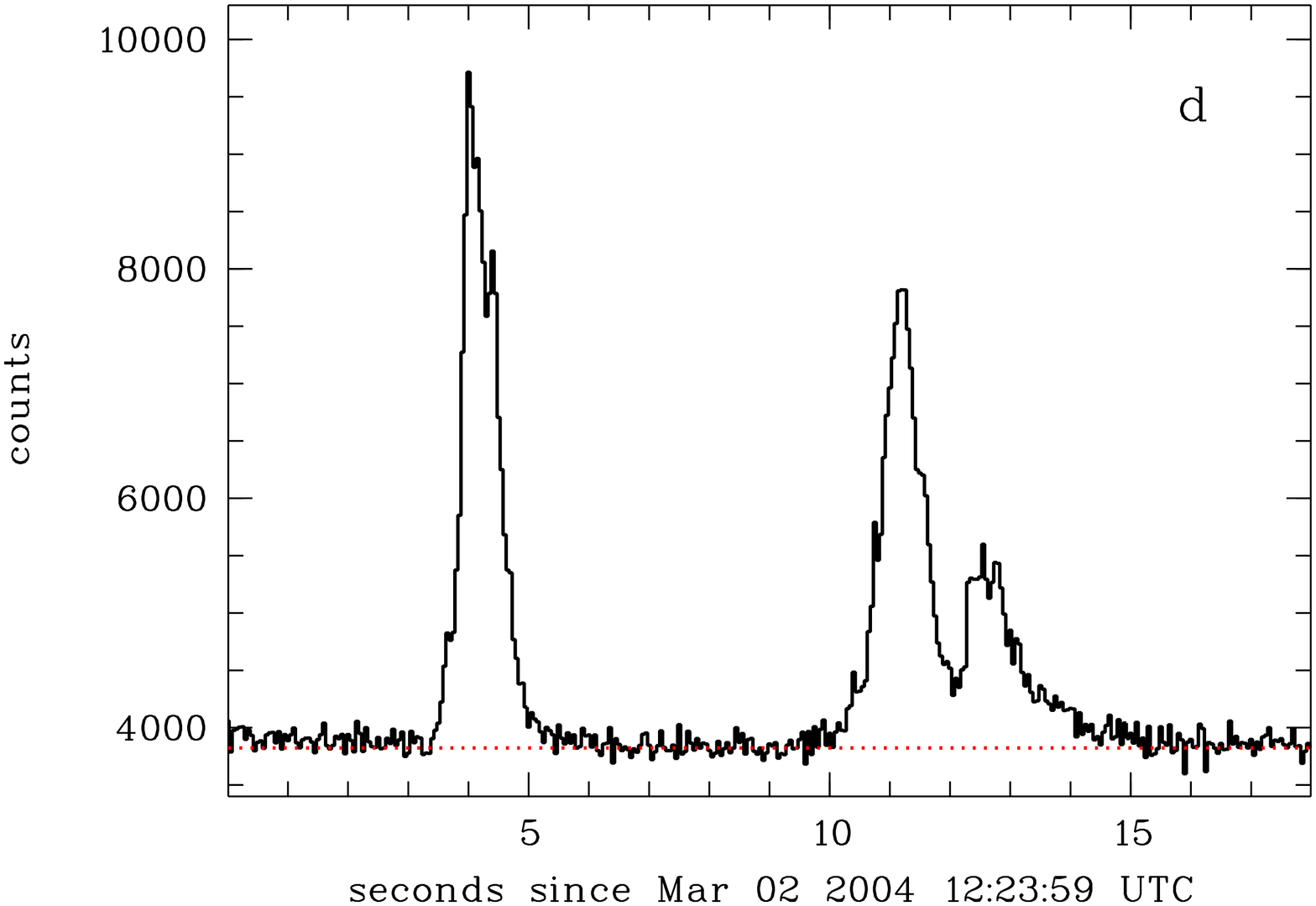}
   \includegraphics[width=0.31\textwidth,angle=270]{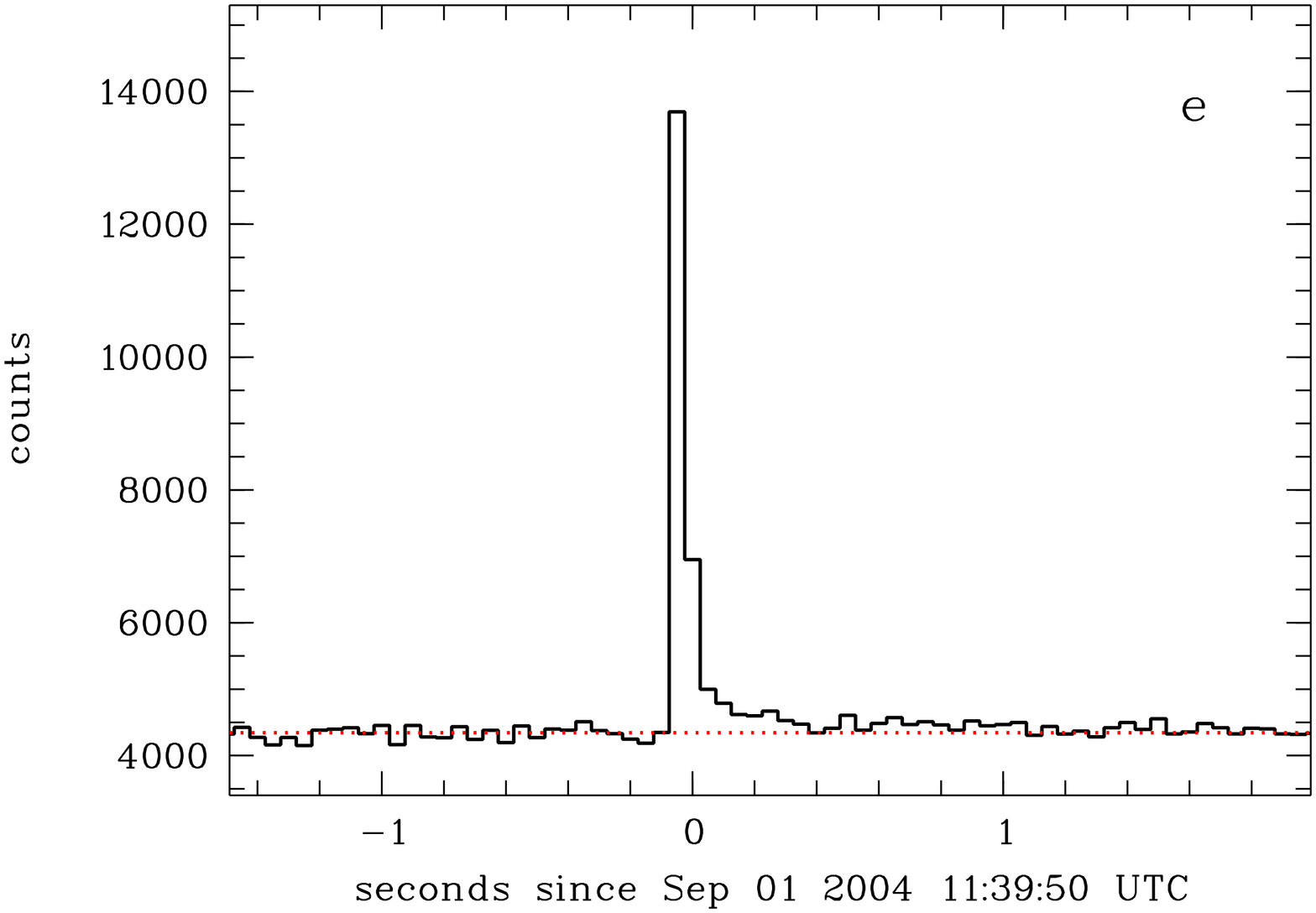}
   \includegraphics[width=0.31\textwidth,angle=270]{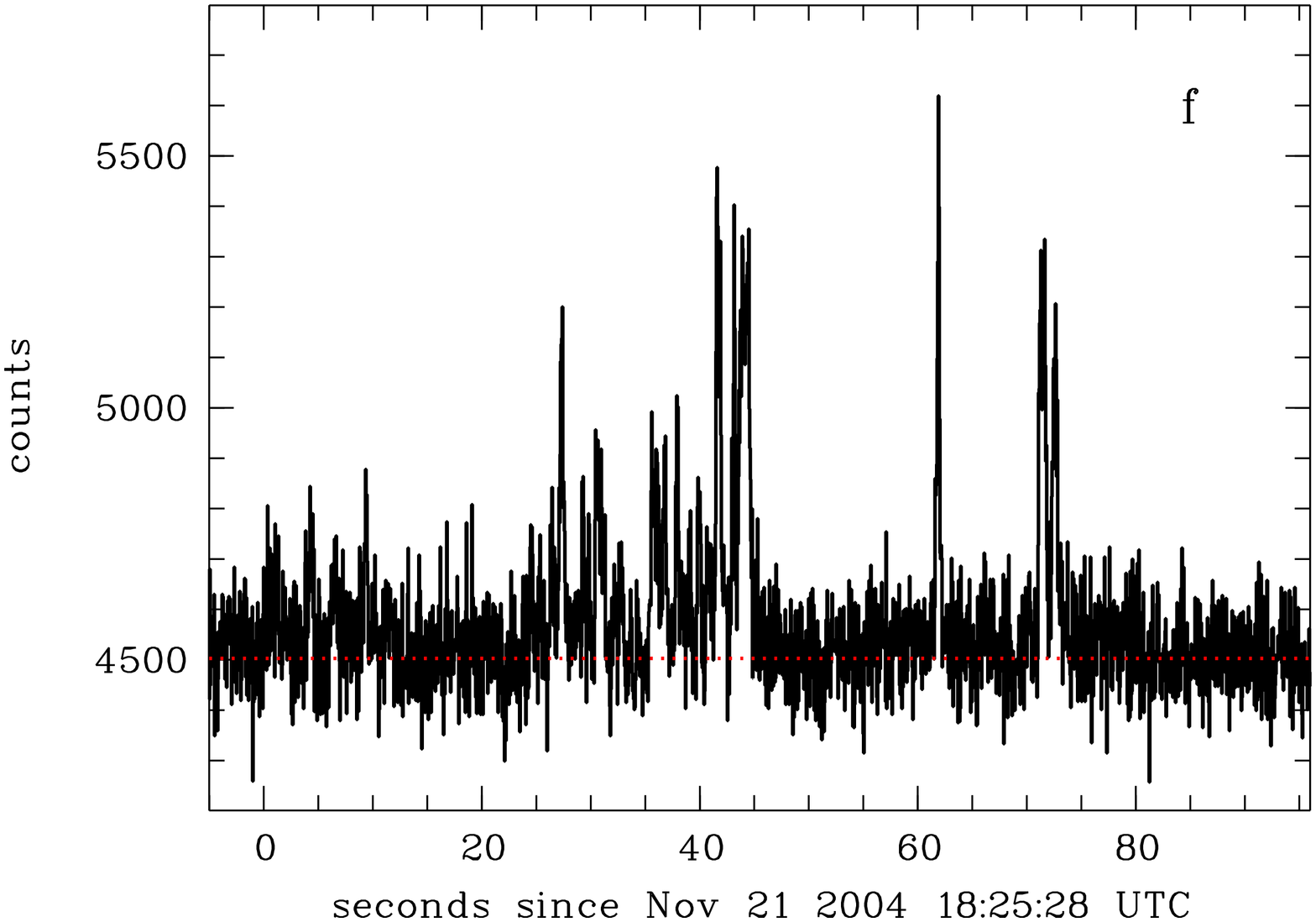}
   \includegraphics[width=0.31\textwidth,angle=270]{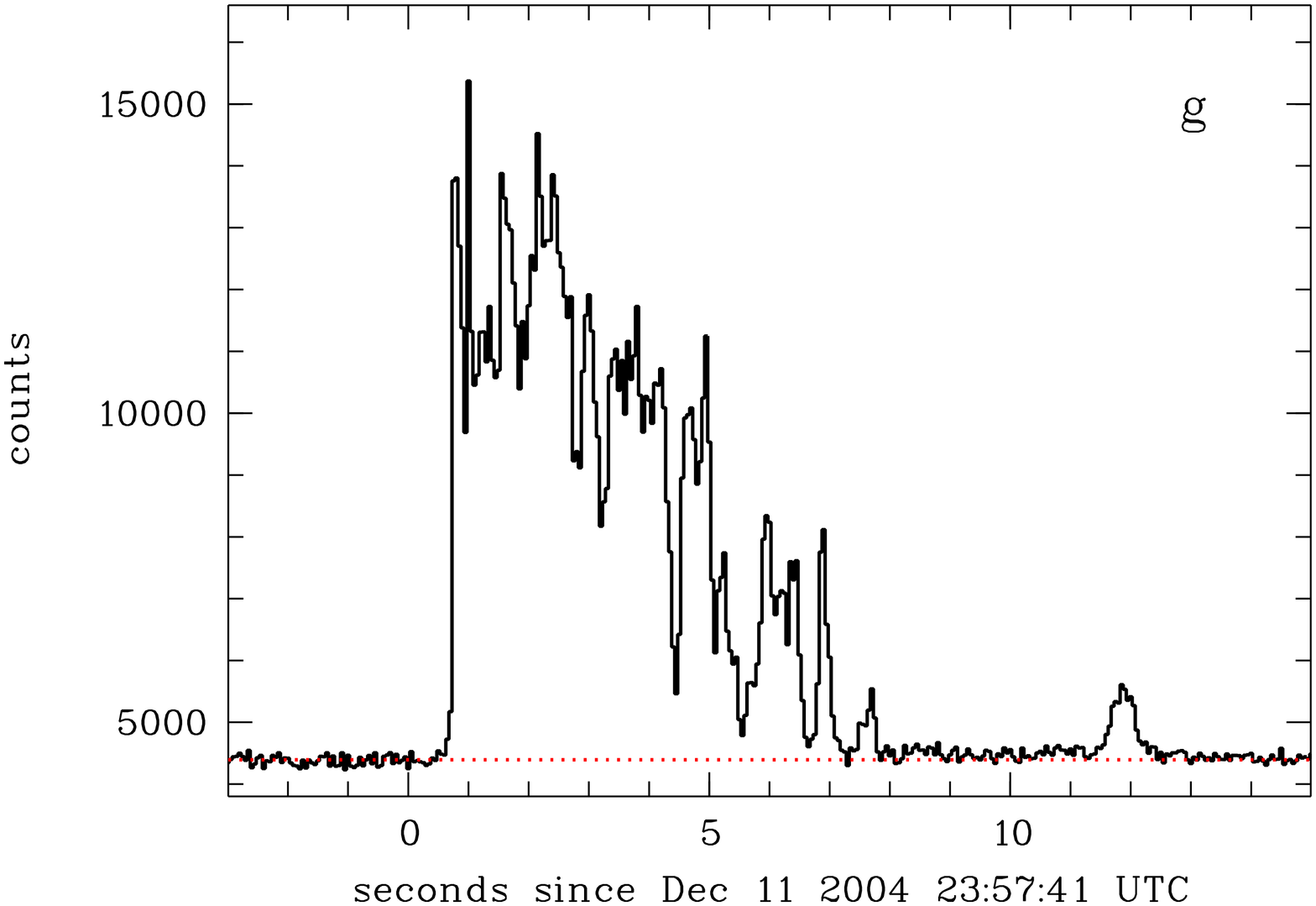}
   \includegraphics[width=0.31\textwidth,angle=270]{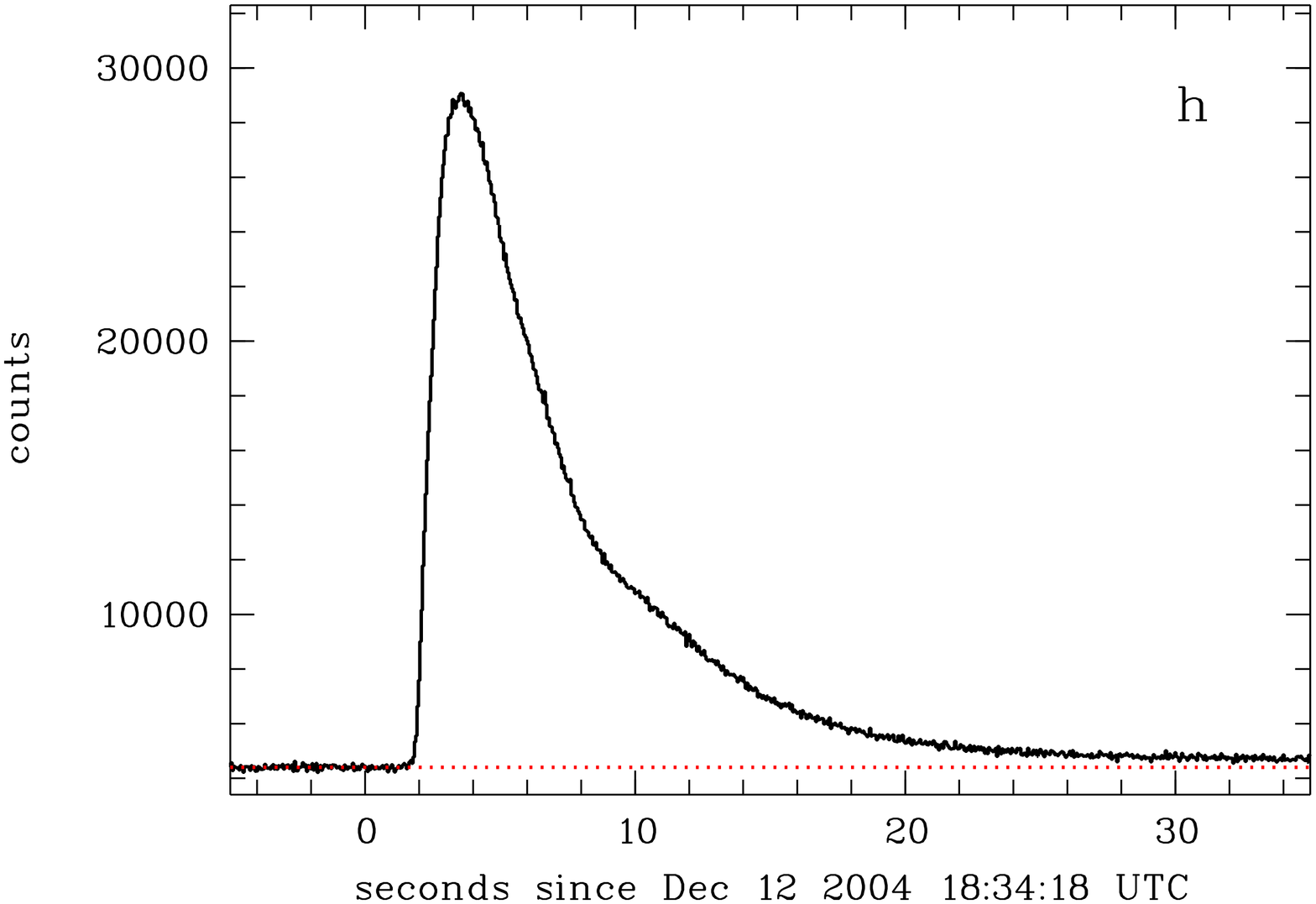}
   \caption{SPI/ACS overall  count rate light curves  with 50\,ms time
   binning  for a  selection of  candidate GRBs  from our  sample. The
   horizontal dotted lines mark  the background level.  {\bf a)} short
   GRB 030711.  Note that  triangulation cannot exclude SGR 0525-66 as
   the origin (Hurley \etal\ 2003a);  {\bf b)} long GRB 030814 (Hurley
   \etal\ 2003b);  {\bf c)} bright short  GRB 031214 with  a very hard
   ($E_P=$2000$\pm$80\,keV) spectrum (Hurley \etal\ 2003c, Golenetskii
   \etal  2003a,2003b); {\bf  d)} triple-peaked  GRB 040302;  {\bf e)}
   bright  unconfirmed event;  {\bf  f)} multi-peaked  GRB 041121;  g)
   structured GRB 041211; h) very bright prototypical FRED (fast rise,
   exponential decay) GRB 041212.}
   \label{fig:lcExamples} \end{figure*}

The  base sample  was constructed  from  all triggered  events in  the
SPI-ACS  overall rate light  curve. Each  trigger in  this preliminary
sample was subsequently checked for particle or solar origin using the
INTEGRAL  Radiation Monitor (IREM)  and the  Geostationary Operational
Environmental   Satellites   (GOES)\footnote{http://www.sec.noaa.gov},
respectively.  Triangulated events with a localization consistent with
the  position of a  Soft Gamma  Repeater (SGR)  were removed  from the
sample.  A significance threshold was  then applied to each event with
a  possible cosmic origin,  such that  a trigger  was included  in the
sample  of  GRB  candidates   when  its  significance  $S$  above  the
background $B$ exceeded $S$=12$\sigma$ in any time interval during the
event. Here, 1$\sigma$  corresponds to 1.57$\times\sqrt{B}$, where the
factor  1.57  takes  into   account  the  measured  deviation  of  the
background from  a Poissonian distribution (von  Kienlin \etal\ 2003b,
Ryde \etal\  2003).  The conservative threshold  of $S$=12$\sigma$ has
been chosen in  order to minimize the contamination  by weak events of
non-GRB origin. Examples  of candidate GRBs from the  final sample are
shown in Fig.~\ref{fig:lcExamples}.

%------------------------------------------------------------------
%------------------------------------------------------------------

Based  on the  selection  criteria  described above,  a  total of  374
candidate  GRBs were  detected  between  Oct. 27,  2002  and Jan.  12,
2005. In  addition, 14 GRBs  with a  significance below  the selection
threshold, but  confirmed by other gamma-ray  missions, were detected.
From  the total  sample  of 388,  a  cosmic origin  for  179 has  been
confirmed.   Most of these  events were  detected by  Konus/{\it Wind}
and/or {\it Ulysses}.  Unfortunately, the {\it Ulysses} GRB experiment
had to be turned off in December 2003 due to the diminishing output of
the radioisotope thermoelectric generator.

A  summary  of  the  currently  active  missions  and  the  confirming
detections    of   SPI-ACS    candidate   GRBs    is    presented   in
Tab.~\ref{tab:confirmations}.   The  number   of  events  detected  by
SPI-ACS and HETE-2  is relatively low compared to  the total number of
GRBs detected by HETE-2 during the INTEGRAL mission time.  This is not
surprising;  it is the  result of  the low  sensitivity of  SPI-ACS to
X-ray flashes  and X-ray  rich bursts, which  constitute approximately
2/3  of the HETE-2  GRB sample  (Lamb \etal  2005).  These  events are
characterized  by an  E$_{p}$$<$80\,keV and  thus have  their emission
maximum below the SPI-ACS energy range.  The number of bursts detected
both in  the INTEGRAL FoV and SPI-ACS  is very low as  well, since the
effective  area of  SPI-ACS is  a minimum  for events  irradiating the
satellite from that direction.

\begin{table}[h]
\caption{Gamma-ray  burst  detectors  and  satellites  and  number  of
  reported  simultaneous  burst detections  with  SPI-ACS. (1): single
  photon counting, (2): 64\,ms resolution during flare mode.  }
\begin{center}
\begin{tabular}{lccc}
\hline\hline
Instrument & Energy & Time Res. & Conf. \\
& [keV] & [s] & \\
\hline
Konus/{\it Wind} & 10--10000 & 0.002--2.39 & 164 \\
{\it Ulysses} & 25--150 & 0.008--2.0 & 92 \\
{\it Mars Odyssey} & 30--8000 & 0.032--0.25 & 79 \\
HETE-2 & 6--400 & $^{(1)}$ & 42 \\
RHESSI & 3--20000 & $^{(1)}$ & 79 \\
INTEGRAL FoV & 3--15000 & $^{(1)}$ & 4 \\
Helicon/{\it Coronas-F} & 10-10000 & $^{(2)}$& 3 \\
{\it Swift} & 15--350 & $^{(1)}$ & 3 \\
RXTE/ASM & 1.5--12 & 100 & 2 \\
\hline\hline
\end{tabular}
\label{tab:confirmations}
\end{center}
\end{table}

Taking  into  account  the  observation  time losses  due  to
radiation  belt  passages  and  regular SPI detector  annealings,
SPI-ACS was  actively observing GRBs  for a total 21.6\,  months up to
Jan.   12,  2005\footnote{The INTEGRAL  satellite  is in  a
highly elliptical  orbit with  a period of  72\,hours.  Due  to the high
particle flux  the instruments are deactivated during  the perigee passage
and  are online  only  for $\sim$62\,hours  per orbit.   Approximately
twice a year, the SPI Germanium detectors   need to be heated up and
cooled  again (annealing) to  counter the  degradation of  the crystal
structure by  the space environment (cosmic rays,  solar wind, etc.)}.
This time  corresponds to  a rate of  216 candidates and  92 confirmed
bursts per  year selected according to our  criteria.  For comparison,
Lichti \etal (2000)  predicted the rate of GRBs prior  to the start of
INTEGRAL to be $\sim$160 per year at a 10\,$\sigma$ level.

\section{Sample Analysis}

For each candidate GRB  in the sample, a linear  fit to the background
in a  time interval before and  after the event was  performed and the
resulting background  counts subtracted.  Next, a  correction for dead
time in the detector and  electronics ($<$3\% for the brightest events
in  the sample\footnote{Note that  the  initial peak  of  the outburst  of
SGR1806-20 from Dec 27 2004,  for which the SPI-ACS data were strongly
affected by dead  time effects, was about 10$^{2}$  times stronger than
the most luminous GRB in the sample.}) was applied. The observer frame
duration T$_{90}$,  the time interval  over which a burst  emits from
5\% to 95\% of its total measured  counts,  was obtained and integrated
burst counts  over the whole event as  well as the peak  counts in the
brightest  0.25\,s  interval  were  estimated.  Due to  the  lack  of
spectral response, these detector counts cannot
be  directly converted  into more  physical units  (fluences  and peak
fluxes).

For a  quantitative evaluation  of the uniformity  of the  SPI-ACS GRB
sample we applied the  $V/V_{max}$ test (Schmidt 1968). In particular,
$V/V_{max}$ compares the volume in which the event was detected to the
largest  possible volume  in which  the event  would be  just  at the
selection  threshold.  The ratio  of the  volumes translates  into the
ratio of the  peak counts, $C_p$, over the  minimum triggering counts,
$C_{min}$,      in      the      same     time      interval      as:
$V/V_{max}=(C_p/C_{min})^{-3/2}$. The time intervals used range from
the  0.05\,s time  resolution  of  SPI-ACS  to  the T$_{90}$  of  the
individual events.

In  order to  estimate the  variability of  the candidate  burst light
curves in the  sample we used the variability  formulation of Reichart
\etal  (2001),  $V_{f=0.45}$.   This  measure  is  computed  from  the
variance  of  the  50\,ms  binned   light  curve  with  respect  to  a
boxcar--smoothed version  of itself.  Following  previous attempts, we
used t$_{45}$ as the smoothing  time scale; this is the effective time
for which  the integrated counts of  the brightest parts  of the event
are 45\% of the total counts.  The deviation of the SPI-ACS background
from a  Poissonian distribution was  taken into account  by broadening
the distribution by a mean factor of 1.57.

All  388 GRBs  and  candidates in  the  SPI-ACS sample  are listed  in
Tab.~\ref{tab:sample} (full  version available  at  the CDS)  together
with the  estimated burst  properties. A continuously  updated on-line
version                                                              is
available\footnote{www.mpe.mpg.de/gamma/science/grb/1ACSburst.html}
including  figures and  ASCII files  of  the light  curves. The  first
column of Tab.~\ref{tab:sample} gives the  number of the event in this
catalogue.   The trigger  date  (YYMMDD)  and UTC  are  listed in  the
following  two  columns.   The  next  four columns  give  the  maximum
significance of  the event above  the background, T$_{90}$,  the total
integrated counts in the event  and the peak counts over 0.25\,s.  The
$V/V_{max}$  statistic and  variability measure  are shown  in columns
eight  and  nine, respectively.   The  final  column  lists all  other
gamma-ray instruments and missions  from which detections of the event
were  reported.   Here, ``U''  stands  for  {\it  Ulysses}, ``K''  for
Konus/{\it  Wind}, ``M'' for  {\it Mars  Odyssey}, ``H''  for HETE-2,
``R''  for  RHESSI,  ``C''  for  Helicon/{\it  Coronas-F},  ``I''  for
INTEGRAL (SPI/IBIS), ``S'' for {\it Swift} and ``X'' for RXTE/ASM.

\begin{table*}[t]
\caption{Excerpt  of  the  SPI-ACS  GRB  sample. Number  of  event  in
catalogue,  date, UTC,  maximum significance  of the  event  above the
background,  T$_{90}$, integrated  counts in  the event,  0.25\,s peak
counts,  variability  and  $V/V_{max}$  are given  together  with  the
confirmations  by other  gamma-ray instruments  (``U'':  {\it Ulysses},
``K'': Konus/{\it  Wind}, ``M'':  {\it Mars Odyssey},  ``H'': HETE-2,
``R'':  for RHESSI,  ``C'': Helicon/{\it  Coronas-F},  ``I'': INTEGRAL
(SPI/IBIS), ``S'': {\it Swift} and ``X'': RXTE/ASM. The complete Table
is available at CDS.}
%\begin{small}
\begin{tabular}{rccccccccc}
\hline \hline
\# & Date & UTC & $\sigma$ & $T_{90}$ & $C_{int}$ & $C_{max}$ & $V/V_{max}$ & $V_{f=0.45}$ & Conf. \\
& & & & [s] & [kcnts] & [kcnts/0.25\,s] & & &  \\
\hline \hline
1 & 021027 & 08:33:51 & 5 &&&&&&  UMK \\
2 & 021102 & 15:58:44 & 30 &7.7$\pm$0.4 &39$\pm$2 & 3.7$\pm$0.2 & 0.25 & 0.22$\pm$0.01 &  UKR \\
3 & 021113 & 13:37:33 & 4 &&&&&&  UKR \\
4 & 021114 & 21:28:54 & 17 &0.1$\pm$0.05 &2.7$\pm$0.2 & 2.7$\pm$0.2 & 0.57 & &  \\
5 & 021115 & 00:44:32 & 14 &0.05$\pm$0.1 &1.7$\pm$0.2 & 1.7$\pm$0.2 & 0.74 & &  \\
6 & 021116 & 08:06:29 & 35 &15$\pm$0.8 &63$\pm$4 & 2.5$\pm$0.2 & 0.20 & 0.15$\pm$0.01 &  \\
7 & 021117 & 23:42:18 & 34 &0.25$\pm$0.05 &7.9$\pm$0.4 & 6.8$\pm$0.2 & 0.21 & &  \\
8 & 021121 & 19:37:20 & 14 &0.05$\pm$0.1 &2.3$\pm$0.2 & 2.3$\pm$0.2 & 0.74 & &  \\
9 & 021125 & 05:59:14 & 34 &20$\pm$2 &66$\pm$6 & 3.5$\pm$0.2 & 0.21 & 0.31$\pm$0.01 &  UK \\
10 & 021125 & 17:58:23 & 34 &19$\pm$2 &68$\pm$5 & 1.4$\pm$0.2 & 0.21 & 0.05$\pm$0.02 &  UMKRI \\
. & . & . & . & . & . & . & . & . & .\\
379 & 041213 & 06:59:36 & 28 &0.05$\pm$0.05 &5.1$\pm$0.3 & 5.1$\pm$0.2 & 0.27 & & KR  \\
380 & 041213 & 08:17:53 & 19 &0.15$\pm$0.05 &3.5$\pm$0.3 & 3.3$\pm$0.2 & 0.49 & &   \\
381 & 041215 & 11:48:55 & 10 & & & & & & K \\
382 & 041219 & 01:42:00 & 78 &130$\pm$9 &360$\pm$30 & 3.9$\pm$0.2 & 0.06 & 0.01$\pm$0.01 &  MKHIXS \\
383 & 041226 & 17:22:26 & 330 &12$\pm$2 &550$\pm$10 & 22$\pm$0.3 & 0.01 & 0.04$\pm$0.01 & MK  \\
384 & 041227 & 18:55:03 & 19 & 27$\pm$2 & 47$\pm$6 & 1.4$\pm$0.2 & 0.48 & & K \\
385 & 041229 & 16:21:00 & 21 &170$\pm$10 &140$\pm$20 & 1$\pm$0.2 & 0.40 & &   \\
386 & 041230 & 06:25:59 & 27 &8.7$\pm$1 &38$\pm$3 & 2.4$\pm$0.2 & 0.28 & &   \\
387 & 041230 & 10:13:40 & 37 &0.05$\pm$0.05 &6.3$\pm$0.3 & 6.4$\pm$0.2 & 0.18 & &   \\
388 & 050112 & 11:10:23 & 177 &0.4$\pm$0.05 &53$\pm$0.8 & 32$\pm$0.3 & 0.02 & &   \\
\\
\hline\hline
\label{tab:sample}
\end{tabular}
%\end{small}
\end{table*}

The table has numerous events with missing entries. For all GRBs which
were confirmed by other instruments but detected by SPI-ACS below the
sample selection  threshold, only the time, date,  significance and common
instuments  are  listed.   Furthermore,  the  variability  measure  was
obtained only for long-duration events with sufficient signal-to-noise
ratio.

\section{Results}

\subsection{Durations}

\begin{figure}[h]
   \centering
   \includegraphics[width=0.345\textwidth,angle=270]{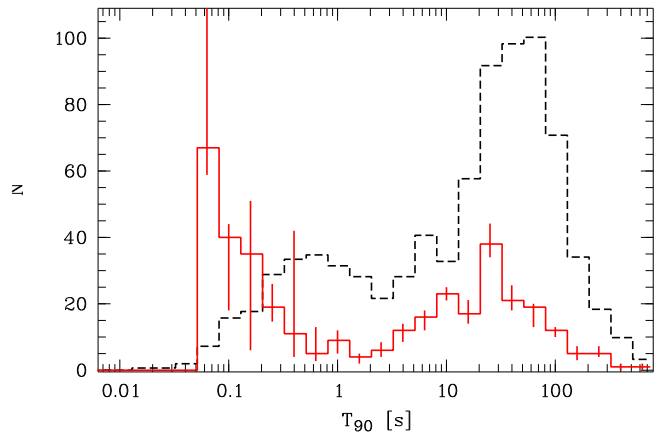}
   \caption {Distribution of T$_{90}$ for all SPI-ACS GRB candidates (solid
line) and for 1234 GRBs from the 4$^{th}$ BATSE GRB catalogue
(Paciesas \etal\ 1999; dashed line) scaled to 21.6\,months of continuous
observation. Note the difference in the distributions of the short events.}
         \label{fig:burstDurationAll}
   \end{figure}

The distribution of  the T$_{90}$ durations is shown in
Fig.~\ref{fig:burstDurationAll}.   For   comparison  the  distribution
derived from 1234  GRBs in the 4$^{th}$  BATSE GRB catalogue
(Paciesas  \etal 1999)  was  scaled  to the  elapsed  mission time  of
SPI-ACS and included in the figure.

%------------------------------------------------------------------
%------------------------------------------------------------------

The distribution  of the SPI-ACS  burst candidates shows  a bimodality
similar to what was observed  by the BATSE experiment.  The long burst
population  has its  maximum at  $\sim$30\,s and  resembles  the BATSE
results closely.  A  short duration component is found  in the SPI-ACS
sample as well,  but it deviates significantly from  the BATSE sample.
While the BATSE  short burst log$T_{90}$ distribution can  be fit by a
Gaussian  centered at  $T_{90}$=0.45\,s (Horvath  1998) the  number of
short burst candidates in the SPI-ACS sample rises steeply towards low
values of $T_{90}$  and has its maximum at the  time resolution of the
instrument   (0.05\,s).     Note   that   number    of   events   with
$T_{90}$$\leq$0.05\,s has to be considered  as a lower limit as events
detected  only  in a  single  time bin  are  rejected  already by  the
software which monitors the telemetry stream (see Sect.~2).

A further difference between the two samples is found in the ratios of
short  to long events.   While the  4$^{th}$ BATSE  catalogue included
$\sim$70\%   long   ($T_{90}>$2\,s)   and  $\sim$30\%   short   bursts
($T_{90}<$2\,s), around 50\% of the total events in the SPI-ACS sample
have $T_{90}$$<$2\,s.  Note that the division into short and long
bursts is not unambiguously defined, as different class boundaries are
generally  found  when   other  properties  (e.g.   fluence,  spectral
hardness  ratios,  $<V/V_{max}>$ are  included  (e.g. Mukherjee  \etal
1998, Hakkila \etal  2000).  In addition, the SPI-ACS  and BATSE burst
samples  have been  compiled  using inherently  different trigger  and
instrumental characteristics which imply different definitions for the
short and long classes in both instruments.

The total number of detected events in SPI-ACS is of the order of 50\%
of the  BATSE detections, when scaled  to a common  time interval.  On
the other  hand, BATSE and  SPI-ACS are sensitive to  different energy
bands. The SPI-ACS lower energy threshold of $\sim80$\,keV reduces the
detection capability for bursts  with peak energies below that energy.
This affects  mainly X-ray flashes ($E_{p}$$<$30\,keV)  and X-ray rich
bursts  ($E_{p}$$\sim$70\,keV).  For  comparison,  BATSE could  detect
events  down  to  $E_{p}$$<$25\,keV   and  was  therefore  capable  of
detecting a significant fraction of the X-ray rich burst population.

Fig.~\ref{fig:burstDurationAll}  indicates  a   slight  shift  of  the
long-duration distribution components  in SPI-ACS towards lower values
of  $T_{90}$ compared to  the BATSE  sample.  Again,  this could  be a
result of the  lower sensitivity at softer energies  for the long (and
generally spectrally soft) distribution. In addition, it is typical to
find a hard-to-soft evolution  during the prompt emission phase.  This
also  leads  to  shorter   durations  at  higher  energies.   Finally,
measurements of  burst properties are  instrument dependent, primarily
due  to siginal-to-noise  limitations  and temporal  binning. This  is
likely  an  additional  reason  for  the difference  in  the  observed
duration distributions.

\begin{figure}[h]
   \centering
   \includegraphics[width=0.35\textwidth,angle=270]{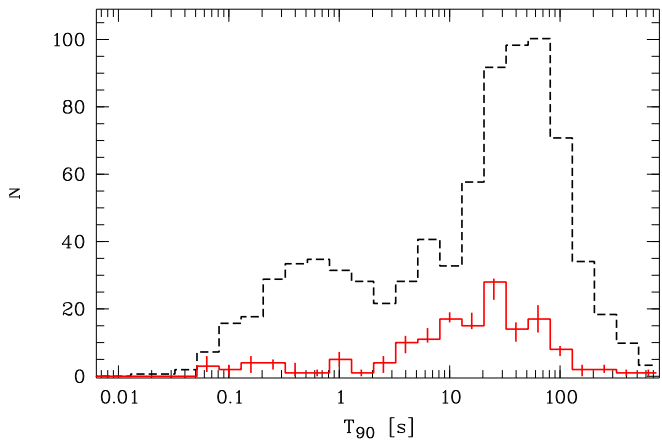}
   \caption { Same as Fig.~\ref{fig:burstDurationAll} except that
     only  the  confirmed  bursts from the  SPI-ACS  sample  are shown.   A
     comparison with  Fig.~\ref{fig:burstDurationAll} reveals that for
     most of  the long-duration bursts  a cosmic origin  was confirmed
     while  this  is true  only  for a  small  fraction  of the  short
     events.}
         \label{fig:burstDurationLoc}
   \end{figure}

The SPI-ACS short-duration population can  be fit by a Gaussian with a
maximum at the time resolution  of 50\,ms.  This shift with respect to
the  BATSE  result is  not  of  instrumental  origin, as  the  SPI-ACS
detectors are  very sensitive  at the typical  peak energies  of short
GRBs.  Also,  a significant hard-to-soft evolution is  not expected to
occur and  cause this  offset.  The deviations  seem to be  mainly for
events   with  $T_{90}$$<$0.25\,s.   The   general  lack   of  spatial
information in the data makes  it difficult to determine the origin of
these events.  In Fig.~\ref{fig:burstDurationLoc} we show the duration
distribution for  the sample of  confirmed GRBs.  While  $\sim$70\% of
the long-duration  bursts are confirmed,  less than 11\% (8\%)  of the
GRB  candidates  with $T_{90}$$<$2\,s  ($<$0.25\,s)  were detected  by
other gamma-ray instruments.

%------------------------------------------------------------------
%------------------------------------------------------------------

Note that  we found  no evidence of  the proposed  third (intermediate
duration) burst  population whose signature  was detected in  the 
BATSE sample (Horvath 1998).  This population was claimed to to
have a softer spectrum than  the long-duration GRB sample.  Due to the
variety of trigger timescales used by SPI-ACS and the lack of spectral
information, it is  not clear how the temporal  sensitivity depends on
the spectral hardness  and peak flux in the  trigger.  Thus, it apears
difficult to  ascertain whether SPI-ACS  is indeed insensitive  to the
intermediate GRBs  or whether some  intermediate GRBs are  included in
the sample  but for some  reason not clearly distinguished  from other
bursts.

\subsection{The Origin of Events with $T_{90}$$<$0.25\,ms}

There are various possible  explanations for the significant excess of
unconfirmed  0.05--0.25\,s  long  events  in the  SPI-ACS  sample  (an
example is shown in  Fig.~\ref{fig:lcExamples}e).  We list them below,
and present arguments for and against each of them.

(i) SPI-ACS is  observing a real population of  short, spectrally very
hard GRBs,  most of  which must have  been undetected by  BATSE.  However, the trigger  efficiency of BATSE was quite  sensitive to hard
photons  (Pendleton \etal\  1998) and  these short  hard  events would
certainly  have been  seen.   Another argument  against  the idea  of
observing  a   real  burst   population  is  that   several  gamma-ray
instruments in orbit  have the capability to detect  such hard events,
as  demonstrated  by the  {\it  Mars  Odyssey},  Konus/{\it Wind}  and
Helicon/{\it Coronas-F}  detection of an  event from December  14 2003
(Fig.~\ref{fig:lcExamples}c);   the  number  of   such  confirmations,
however, is too low to explain the excess of very short events in this
way.

(ii) A  small contribution  to the short  burst population  might come
from outbursts  of Soft Gamma  Repeaters.  These objects  are strongly
magnetized neutron stars, ( ``magnetars'' -- Duncan \& Thompson 1992),
which are characterized by brief (typical $\sim$100\,ms), very intense
(up   to   10$^{44-45}$\,erg)  bursts   of   hard   X-rays  and   soft
gamma-rays.  Their light  curves  resemble those  of  the short  GRBs.
Thus, events in the SPI-ACS count rate originating from known SGRs are
not easily distinguishable from short GRBs without the localization by
triangulation with  other instruments.  However, the  vast majority of
these bursts  have soft spectra and  most of them are  not detected by
the SPI-ACS.

(iii)  The  evaporation  of  primordial black  holes  through  Hawking
radiation  can lead  to a  sudden burst  of gamma-rays  with durations
$\sim$100\,ms (Halzen  \etal 1991). This has been proposed for the
BATSE  sample  of  very  short   bursts  by  Cline  \etal  (2003).  But
the spatial distribution of those events ($<$$V/V_{max}$$>$$\sim$0.76)
is not consistent with ours.  Our
sample  of events  with  $T_{90}$$<$100\,ms is  consistent with
isotropy in a Euclidean geometry, $<$$V/V_{max}$$>$$\sim$0.48.

\begin{figure}[h]
   \centering
   \includegraphics[width=0.5\textwidth]{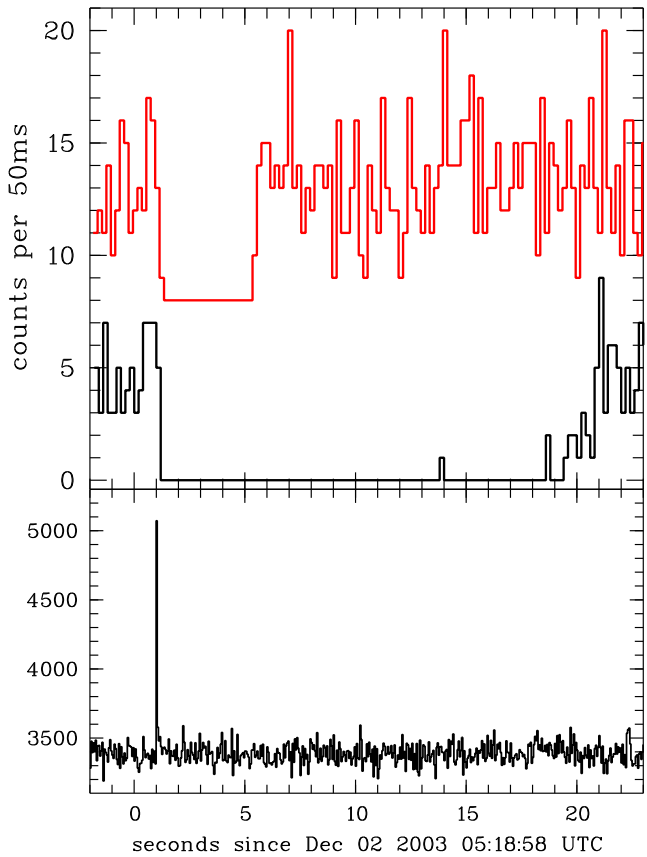}
   \caption {Example of a very short event in SPI-ACS (bottom panel)
     with simultaneous saturation of two SPI Ge detectors (top panel). The
     SPI-ACS light curve is shown with 50\,ms binning and the SPI
     light curves of detector \#9 (upper curve) and \#10 (lower curve)
     are binned with 200\,ms time resolution.}
   \label{fig:1431}
   \end{figure}

(iv) A significant contribution to the short candidates could arise from
instrumental effects and/or cosmic ray events.  Analyzing the data of very
short events together with simultaneous data from the spectrometer SPI
revealed   that  a   significant  fraction of them are  accompanied   by  the
simultaneous saturation of one or several Ge-detectors. An example for
an event on Dec. 1, 2003  is shown in Fig.~\ref{fig:1431}. At the time
of  the significant  rate increase  in SPI-ACS  over  $\sim$50\,ms two
neighboring Ge-detectors,  \#9 and \#10, went into  a saturation which
lasted for $\sim$4 and $\sim$20\,s, respectively.  The analysis of SPI
data from  a subset of  satellite revolutions shows that  a saturation
event occurs  approximately every  four hours and  that nearly  all of
these  events  have a  simultaneous  rate  increase  in SPI-ACS.   The
opposite  approach, an  independent  search for  short  events in  the
SPI-ACS  overall  count rate,  revealed  $\sim$15  events  per hour  at  a
4.5$\sigma$ level.   Combining these results  shows that approximately
every 60$^{th}$  very short SPI-ACS event coincides  with a saturation
in the Ge detectors.

%------------------------------------------------------------------
%------------------------------------------------------------------

The saturation of the Ge detectors  can be explained by the deposit of
a  large amount of energy in  the  crystals.   For example, a  very  energetic
cosmic ray  particle can traverse  a BGO  crystal and  deposit part  of its
energy, thus producing the short count rate increase. Depending on the
flight  direction,  the  particle  can  go on to traverse  one  or  several
Ge detectors and cause the saturation,  or pass through the BGO shield
a second time,  or not  interact with any of the detectors  again.  The
probability of a particle hitting  the SPI-ACS and Ge detectors can be
estimated from the geometry of  the instrument to be $\sim$1/40 of the
probability to pass only  through the anti-coincidence shield. This is
in rough agreement with the rate of SPI-ACS short events that coincide
with saturations.   Therefore, this simple hypothesis  suggests that a
significant  fraction of  the unconfirmed  short burst  candidates can
originate  from  cosmic ray   particles  hitting  the  instrument.
%By
%comparing the  rate of events  with the known spectrum  of cosmic rays
%(Cronin \etal 1997),  the typical energy for a  single particle can be
%estimated to be $\geq$300\,GeV.

\begin{figure}[h]
   \centering
   \includegraphics[width=0.35\textwidth,angle=-90]{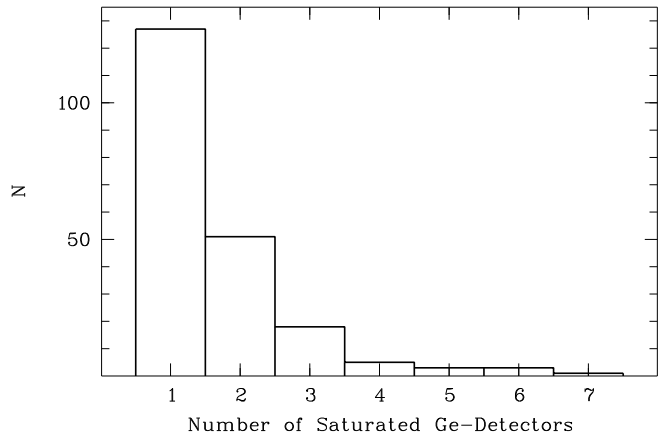}
   \caption {Distribution of saturated Ge detectors of SPI for a
   duration of 15\,days (INTEGRAL mission revolutions 137-141). Most
   saturations occur in a single detector only.
   \label{fig:detectorDistribution}}
   \end{figure}

Most of the  saturation events occur in a  single Ge-detector. Hits in
several     detectors      are     significantly     less     frequent
(Fig.~\ref{fig:detectorDistribution}) but happen either in neighboring
detectors or along a line of detectors. If the saturation is caused by
a single particle then a  particle passing through the coded mask will
only deposit energy  in one detector as the  cross-section for hitting
more  than one  Ge-detector is  low. A  particle hitting  the detector
plane  from the  side can  deposit  energy (and  cause saturation)  in
several  detectors. This  is in  agreement with  the  observation that
saturation events in more than  one detector are always accompanied by
a significant rate increase in  the SPI-ACS, while a small fraction of the
saturation events in single  detectors show no corresponding signal in
the SPI-ACS overall rate.

%------------------------------------------------------------------
%------------------------------------------------------------------

In the  case of a particle  shower, the number of  detectors being hit
depends on  the width of  the distribution of the  secondary particles
and  on   the  inclination   angle.   In  IBIS/PICsIT   strong  bursts
(2500\,counts)  up  to  170\,ms long of  particle-induced  showers  are
observed (Segreto 2002).  These  showers may be produced by cosmic ray
hits on the satellite or  detector structure.  As a consequence of the
interaction of  the primary particle with  matter, high-energy photons
and  electron-positron pairs  are  created. These  in  turn produce  a
cascade of secondary electrons and photons via bremsstrahlung and pair
production. The  time profiles  of the events  observed in  PICsIT are
rather complex and  both straight tracks and closed  areas are visible
(Segreto 2002).   These observations are similar to  the detections in
SPI and thus support the idea of cosmic rays as a partial origin of
the very short SPI-ACS events.

The  observed  rate  increase  in  SPI-ACS  can  be  produced  by  the
deceleration of a cosmic ray particle in the BGO crystal. This induces
a long-lasting (50-150\,ms)  phosphorescent afterglow energetic enough
to cause  recurrent triggers in  the electronics.  BGO  was originally
selected for  SPI-ACS because  of its short  decay times and  very low
phosphorescence compared  to e.g. CsI(Na) and NaI(Tl).   BGO has decay
times  of 60\,ns  and 300\,ns  where  the second  is by  far the  more
probable (90\% vs.   10\% for the 60\,ns) and  an afterglow of 0.005\%
is expected at 3\,ms (Farukhi 1982).

In order to  produce a single count in the SPI-ACS  rate, a minimum of
80\,keV  has   to  be  accumulated  during  an   integration  time  of
600\,ns. With a  total light yield for BGO  of 8000-10000 photons/MeV,
this corresponds  to $\sim$675 photons.  Assuming  an exponential decay,
the  radiation will decrease to 0.005\%  at 3\,$\mu$s. Thus
the decay is so fast that only a small number of counts ($\sim$3) will
be  produced.   From the  afterglow  properties  the minimum  original
excitation  energy  can be  estimated.   For  instance,  to have  1000
recurrent  triggers  (a count  rate  increase  of  1000) in  a  single
crystal, the afterglow  must be bright enough to  be above the 80\,keV
threshold for  $\sim$3\,ms (integration time + dead  time).  Thus, the
0.005\,\% afterglow emission at  3\,ms corresponds to a $\sim$1.6\,GeV
initial  excitation  using the  light  yield  given above.  Therefore,
particles which deposit $\sim$1.6\,GeV or more in a crystal can indeed
be the origin  of the short event population in the SPI-ACS. Short events
showing similar  temporal behavior have been discussed for CsI(Na)
and  CsI(T1)  crystals exposed  to  primary  cosmic radiation  (Hurley
1978). The same conclusion was reached for the origin, namely cosmic ray nuclei in
the  iron group. Note  that CsI  has a  significantly more
intense afterglow (0.1--1\,\% at 6\,ms).

We  performed simulations  of  particle attenuation  in the  detectors
using  {\it SRIM}\footnote{http://www.srim.org/}.  The  results showed
that heavy cosmic ray  nuclei (e.g., Fe) with energies  of several GeV
per ion can easily produce the observed effects in the SPI-ACS and SPI.

\subsection{Intensity}

\begin{figure}[h]
   \centering
   \includegraphics[width=0.35\textwidth,angle=270]{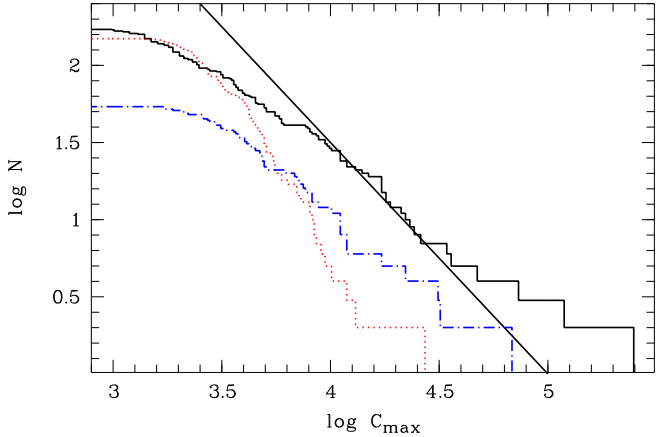}
   \caption  {log$N$--log$C_{max}$ distribution for  three sub-samples,
   together  with  the  expected  power  law  slope  of  -3/2  for  a
   homogeneous   distribution   (solid   line).    The   long duration
   ($T_{90}$$>$2.5\,s,   solid  line)   and   short duration
   (0.25\,s$<$$T_{90}$$<$2.5\,s, dash-dotted line) samples follow a smiliar
   power  law with  a  slope  of $\sim$-1.1.   The  very short  events
   ($T_{90}$$<$0.25\,s, dotted) have a significantly steeper distribution.}
         \label{fig:logNlogPCnts}
   \end{figure}

The distribution of burst intensities can provide valuable information
on the  radial distribution of  the sources.  The  integral brightness
distribution is  expected to follow a  power law with a  slope of -3/2
for  a  homogeneous  distribution  of  sources  assuming  a  Euclidean
geometry. The lack of spectral  resolution limits the analysis for the
SPI-ACS  sample to  detector count  units. Fig.~\ref{fig:logNlogPCnts}
shows the integral  log$N$--log$C_{max}$ distributions for peak counts
$C_{max}$ of  three sub-samples measured over a  timescale of 0.25\,s.
The solid, dashed and dotted distributions represent the long duration
sample  ($>$2.5\,s), the short  bursts (0.25\,s$\leq$$T_{90}$$<$2.5\,s)
and the  population of  very short events  ($<$0.25\,s), respectively.
The deviation from a --3/2 power  law (solid line) is visible for each
of  the sub-samples.  Both the  long duration  and the  short duration
bursts follow a  $\sim$-1.1 slope. The population of  very short burst
behaves  very differently, with  a significantly  steeper distribution
(slope$\sim$-2.3).    The  turnover   at  low   values   of  $C_{max}$
corresponds to the selection threshold of the sample.

%------------------------------------------------------------------
%------------------------------------------------------------------

A deviation  from a homogeneous  distribution was also found  by BATSE
(Fishman \etal  1994).  Nevertheless, the substitution  of peak counts
for peak fluxes causes an  unknown uncertainty in the SPI-ACS results.
While  in a  first-order approach  each incoming  photon  produces one
detector  count,  the  uncertain  effective  area  (depending  on  the
incident angle) prevents a direct comparison.  Thus, $C_{max}$ depends
on the spectral shape of  the individual events.  Still, a qualitative
statement on the behavior of  the three sub-samples can be made.  Both
short-  and long  duration  bursts  follow the  same  slope while  the
distribution  of the  very  short events  differs  strongly.  This  is
consistent with the  hypothesis that the very short  events are not of
cosmological origin.

\begin{figure}[h]
   \centering
   \includegraphics[width=0.35\textwidth,angle=270]{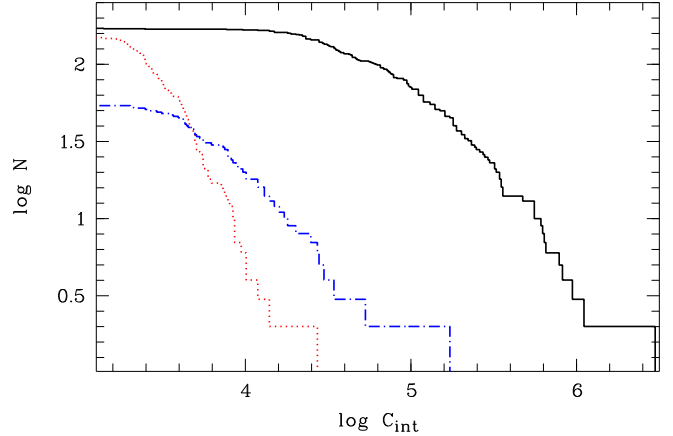}
   \caption {Same as Fig.~\ref{fig:logNlogPCnts} for log$N$--log$C_{int}$. }
         \label{fig:logNlogCnts}
   \end{figure}

Fig.~\ref{fig:logNlogCnts}        shows        the       corresponding
log$N$--log$C_{int}$  distribution.   Here,  the  integrated  detector
counts were used  as a measure of the fluences.  A behavior similar to
the sub-samples in the  log$N$-log$C_{max}$ distribution is found. The
long- and short duration populations evolve similarly while the sample
of very short events displays a significantly steeper distribution.

%------------------------------------------------------------------
%------------------------------------------------------------------

\subsection{$V/V_{max}$}

%------------------------------------------------------------------
%  \begin{figure}[h]
%   \centering
%   \includegraphics[width=0.35\textwidth,angle=270]{hist_VVmax.ps}
%   \caption {}
%   \end{figure}
%------------------------------------------------------------------

Another  test for  homogeneity  is the  $V/V_{max}$  test (Schmidt
1968). If the  burst population is drawn from  a uniform distribution
in Euclidean space, the mean  value of $V/V_{max}$ will be 0.5. For
the individual  sub-samples of  long, short and  very short  events we
find $<$$V/V_{max}$$>$  values of  0.26, 0.31 and  0.48, respectively.
As already  seen in  the intensity distributions,  the short  and long
burst   populations   behave  similarly   and   show  evidence   for
inhomogeneity. This is consistent with the BATSE result, namely
$<$$V/V_{max}$$>$$\sim$0.34   (Fishman  \etal  1994).    In  contrast,
$<$$V/V_{max}$$>$  for  the very  short  events  is  consistent with  a
homogeneous distribution.   Again, this  points to a  different origin
for  them compared  to  the   short-  and
long duration bursts.

\subsection{Variability}

\begin{figure}[h]
   \centering
   \includegraphics[width=0.35\textwidth,angle=270]{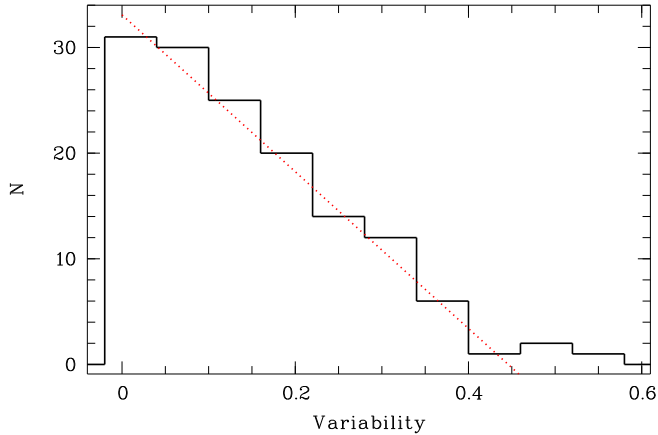}
   \caption {Distribution of the variability measure for 143 candidate bursts
   from the sample (solid line) together with best linear
   fit (dotted).}
   \label{fig:var}
   \end{figure}

In  Fig.~\ref{fig:var} we show  the distribution  of the  time profile
variability  of 143  candidate bursts  using the  measure $V_{f=0.45}$
presented by  Reichart \etal\ (2001).  The number  of bursts decreases
linearly       with       increasing       variability       following
$N$=(-74$\pm$3)$\times$$V_{f=0.45}$+(33$\pm$0.8).    The   empirically
found  correlation   between  $V_{f=0.45}$  and   the  isotropic  peak
luminosity, $L_{iso}$ by Reichart \etal\ suggests that the majority of
the events in the sample are  intrinsically faint and only a few bursts
have $L_{iso}$$>$10$^{53}$\,erg.

%------------------------------------------------------------------
%------------------------------------------------------------------

%------------------------------------------------------------------
%  \begin{figure}[h]
%   \centering
%   \includegraphics[width=0.32\textwidth,angle=270]{hist_intCnts.ps}
%   \caption {}
%   \end{figure}
%------------------------------------------------------------------
%------------------------------------------------------------------
%  \begin{figure}[h]
%   \centering
%   \includegraphics[width=0.32\textwidth,angle=270]{hist_pCnts.ps}
%   \caption {}
%   \end{figure}
%------------------------------------------------------------------

% --------------------------------------------------------------------------
\section{Conclusion}

The anti-coincidence shield of  the INTEGRAL spectrometer SPI is
operating successfully as  an omnidirectional gamma-ray burst detector
above  $\sim$80\,keV.  The 1$^{st}$  catalogue includes  388 candidate
GRBs detected  during the first 26.5\,months of  mission operation.  For
179 triggers a  cosmic origin could be confirmed  by observations with
other  gamma-ray instruments  in  space. The  sample  shows the  known
duration bimodality with a strong excess  at the very short end of the
distribution  ($<$0.25\,s). The  origin of  this population  of events
($\sim$40\% of the total  number) was demonstrated to be significantly
different  from  the   normal  GRB  sample,  as  shown   both  by  the
log$N$--log$C_{max}$ distribution as well  as by the $V/V_{max}$ test.
Observations  of  simultaneous  saturations  in  the  spectrometer  Ge
detectors and very short events  in the SPI-ACS overall rate suggest a
cosmic ray origin for a significant fraction of these events.

The  short-   and  long duration   sample  includes  236   GRBs,  which
corresponds  to  a  detection  rate  of $\sim$130  GRBs  per  year  of
continuous  observation  ($\sim$1 every  three  days).  The  intensity
distribution of this  sample is consistent with the  BATSE results and
shows  a deviation  from  a homogenous  distribution  in Euclidean
space.

% --------------------------------------------------------------------------
\section{Acknowledgments}
We are grateful  for the support from the  ISDC shift team, especially
to J.~Burkowski and  M.~Beck.  We thank the referee,  J.  Hakkila, for
his  insightful  comments. The  SPI-ACS  is  supported  by the  German
``Ministerium  f\"ur Bildung  and  Forschung'' through  the DLR  grant
50.OG.9503.0.   KH is  grateful for  support under  the  INTEGRAL U.S.
Guest Investigator program, NASA grants NAG5-12614 and NNG04GM50G.  We
are also grateful to V. Pal'shin for assistance in calibrating the ACS
timing.

% --------------------------------------------------------------------------

\onecolumn

\twocolumn

\end{document}